\documentclass[journal]{IEEEtran}
\usepackage{amsmath}
\usepackage{amssymb}
\usepackage{color,soul}
\usepackage[pdftex]{graphicx}
\allowdisplaybreaks[4]
\usepackage[noadjust]{cite}

\usepackage{float}
\usepackage{comment}

\usepackage{pgfplots}
  \pgfplotsset{compat=newest}
  \usetikzlibrary{plotmarks}
  \usetikzlibrary{arrows.meta}
  \usepgfplotslibrary{patchplots}
  \usepackage{grffile}
  \usepackage{url}
\usepackage[nolist,printonlyused]{acronym}
\usepackage{listings}
\newfloat{lstfloat}{t}{lop}
\floatname{lstfloat}{Code Block}

\usepackage{algorithm}
\usepackage{algorithmic}
\usepackage{url}
 
\begin{document}

  \title{\texttt{LITS.jl}---An Open-Source Julia based Simulation Toolbox for Low-Inertia Power Systems}

\author{
Rodrigo Henriquez-Auba$^\star$, Jose D. Lara$^\dagger$, Ciaran Roberts$^{\star}$, Nathan Pallo$^\star$ and  Duncan S. Callaway$^{\star, \dagger}$%
\thanks{This research is supported by the National Science Foundation under grants CPS-16446612 and CyberSEES-1539585. This work was supported by the  Laboratory Directed Research and Development (LDRD) at the National Renewable Energy Laboratory (NREL). Corresponding authors: rhenriquez@berkeley.edu and jdlara@berkeley.edu}
\thanks{$^\star$Department of Electrical Engineering and Computer Sciences, University of California, Berkeley, United States}
\thanks{$^\dagger$Energy and Resources Group, University of California, Berkeley, United States}}


\maketitle

\begin{abstract}
The integration of converter-interfaced generation (CIG) from renewable energy sources poses challenges to the stability and transient behavior of electric power systems.  Understanding the dynamic behavior of low-inertia power systems is critical to addressing these stability questions. However, there is a limited availability of open source tools explicitly geared towards low-inertia systems modeling. In this paper, we develop an open source simulation toolbox to study transient responses under high penetration scenarios. \texttt{LITS.jl} is implemented in the Julia computing language and features multi-machine modeling capability, a rich library of synchronous generators components (AVR, PSS, Governor, etc.) and inverter configurations. Several case studies are conducted and benchmarked and validated against existing toolboxes. Case studies are selected to showcase the integration of different device models' behavior in power systems.
\end{abstract}


\begin{acronym}[PSCC]
\acro{AVR}{Automatic Voltage Regulators}
\acro{BES}{Battery Energy Systems}
\acro{CIG}{Converter-Interfaced Generation}
\acro{DAE}{Differential-Algebraic Equations}
\acro{PSS}{Power System Stabilizers}
\acro{PLL}{Phase-Locked-Loop}
\acro{RES}{Renewable Energy Sources}
\acro{ROCOF}{Rate of Change of Frequency}
\acro{SRF}{Synchronous Reference Frame}
\acro{VSC}{Voltage Source Converter}
\acro{VSM}{Virtual Synchronous Machine}
\acro{EMT}{Electromagnetic Transient}
\acro{ERCOT}{Electric Reliability Council of Texas}
\end{acronym}

\section{Introduction}

The increasing penetration of \ac{CIG} presents new challenges to the operation of power systems. The integration of more \ac{CIG} and subsequent substitution of traditional synchronous machines decreases systems' inertia. This may result in higher frequency rate of change, and lower nadir when a disturbance occurs. Theoretical results about the reduction of available inertia in systems have been studied in detail, and raise concerns about the transient stability behavior as well as the potential control actions to mitigate risks after a contingency. An additional complexity introduced by \ac{CIG} are new control structures that have faster dynamics which can interact and generate unexpected time coupling with existing system components - for example, synchronous generators' controls or transmission lines \cite{milano2018foundations, ulbig2014impact}.

The study of stability and analysis of transient response in large-scale low-inertia power systems requires understanding the dynamic behavior of system components under myriad assumptions. These include model fidelity, converter configurations, line dynamics, and load models.  Traditional modeling assumptions, such as omitting network dynamics due to time scale separation, likely break down with  fast acting \ac{CIG} \cite{markovic2018stability,gross2019effect,Standard1110}. To comprehend the effects of increased \ac{CIG} sources on power systems,  it is critical to model  diverse power systems component combinations. However, research concerning the level of modeling detail required to produce relevant conclusions is still ongoing, rendering the development and performance of computational studies cumbersome. 

Practitioners are already beginning to experience the limitations of existing modeling packages for systems that have sizeable \ac{CIG} penetration. For instance, \ac{ERCOT} issued a report \cite{ercot_stability} pointing to non-convergence and numerical instabilities under low system strength conditions. The report also notes the inability to model the specific component behavior during transients. This can have a significant impact on the response of \ac{CIG}, since it does not characterize converter misoperation from erroneous measurements following a disturbance \cite{nerc_bluecut}. 

This paper is rooted in the principle that large scale low inertia power systems require a simulation tool whose flexibility enables researchers to quickly explore emerging challenges. It needs to be open-source so that research scholars from all communities can access it and so that it can undergo rapid research-driven evolution with the introduction of new technologies and as our understanding of the dynamics of these systems improves. Although there are several non-commercial tools available for electro-mechanical modeling in power systems, they do not fully address the needs for open source computing to study large-scale low inertia systems. For instance,  PSAT and Matdyn are based on the commercial software Matlab. The former has a version based on GNU Octave \cite{milano2005psat, cole2010matdyn}. However, the level of customizability in these tools is limited: they do not feature component-level flexibility such as the specification of composite devices like inverters. Meanwhile, PYPOWER-Dynamics and InterPSS are non-commercial alternatives to perform transient simulations based on open source programming languages \cite{pypowergithub, zhou2017interpss}. However, they offer only basic converter models and feature strictly algebraic network models. Finally, DOME is a Unix-supported power system simulation project written in Python, C, and Fortran which has both AC and DC devices and can mix phasor and \ac{EMT} models \cite{milano2013python}. Although DOME is developed using open-source languages and operating systems, it is not an open-source project. The source code is only provided to researchers upon request \cite{dome}.

This paper presents the \textbf{L}ow \textbf{I}nertia \textbf{T}ransient \textbf{S}imulation toolbox \texttt{LITS.jl}, with four key contributions:
\begin{itemize}
    \item \texttt{LITS.jl} is designed to enable researchers to easily assess the trade-off between model complexity and computational requirements.  This enables low-effort comparisons of  system response for different device models under the same disturbance, which is not straightforward with existing tools.
    \item \texttt{LITS.jl} is developed in the Julia computing language on top of the existing package DifferentialEquations.jl \cite{bezanson2017julia, rackauckas2017differentialequations}. These state of the art open-source computing tools are freely accessible to the research community, and have multiple features to facilitate the development of computational experiments.  
    \item \texttt{LITS.jl} separates modeling from simulation, and can quickly scale to analyze large-scale systems by exploiting Julia's large library of solution methods.  Leveraging these libraries enables many benefits, including substantial speed-up on computing time.  
    \item The modular design enables code and model reuse; this reduces development requirements and enables fast and simple prototyping of controls and models. 
\end{itemize}	 

Our choice of Julia as the software development environment is central to enabling many of the features of \texttt{LITS.jl}. Julia is a scripting language like Python and Matlab, but offers the performance one would associate with low-level compiled languages.  It was developed with mathematical and scientific computing in mind and has a number of powerful features that are well-suited to exploring large-scale computing experiments.  Julia's support of multiple dispatch and composition allows us to design a software and model library that is computationally efficient yet easy for a developer to use and extend \cite{methodsjulia}. Multiple dispatch -- which we define precisely in the following section -- allows us to create generic methods for modeling power system devices; this results in a fast, feature-rich, and highly extensible modeling package. The detailed code implementation is discussed in Section~\ref{sec:software}. 

The paper is structured as follows: Section \ref{sec:software} describes the modeling strategy and software architecture, as well as the implementation of \texttt{LITS.jl}. Section \ref{sec:modeling} showcases the different models available and specifies the dynamic system. Section \ref{sec:simulations} contains simulations to highlight the features available in the software. Finally, Section \ref{sec:conclusion} outlines  conclusions and future work.

\section{Software Implementation \label{sec:software}}

\texttt{LITS.jl} uses Julia data structures to define power system dynamic devices and their corresponding components. The components are built exploiting \texttt{PowerSystems.jl} type hierarchy \cite{lara2018power}. The key feature of \texttt{LITS.jl} design is the capability to mix and match different component models with a small number of script modifications. The flexibility of model specification enables precise control when specifying the complexity of each power system device. 

Multiple dispatch lies at the core of Julia's power as a scientific scripting and computing language. It enables the compiler to determine which method to call based solely on the types of the argument passed into a generic function. This is a way to enable code reuse and the implementation of generic interfaces for custom models without requiring modifications to the main source code. For example, \texttt{LITS.jl} will deploy different order generator models by changing by detecting the signature of arguments passed into the model function. Multiple dispatch is particularly useful for mathematical modeling since methods can be defined based on abstract data structures to enable code re-use and easy interfacing with custom models \cite{bezanson2017julia}. Internally, the choice of the mathematical description for a device is implemented as a function method and called based on arguments given and their types. This approach is different from traditional scripting languages, where dispatch is based on a special argument syntax and is sometimes implied rather than explicitly written.

The flexibility in \texttt{LITS.jl} originates from the use of \textit{polymorphism} and encapsulation in the device model definition, which enables treating objects of different types similarly if they provide common interfaces. Generators and inverters are defined as a composition of components; the functionalities of the component instances determine the device model providing a flat interface to the complex object. This design enables the interoperability of components within a generic device definition. As a result, it is possible to implement custom component models and interface them with other existing models with minimal effort. For instance, it is possible to formulate any of the classical machine models in terms of its components and also formulate custom machine models, as shown in the forthcoming section. 

\subsection{Data Structures}

We define a meta-model to implement the generic composition scheme for devices that inject current into the system. Figures  \ref{fig:gen_metamodel} and \ref{fig:inv_metamodel} showcase the generator and inverter meta-model; each block corresponds to a possible component instance. This implementation enables the developer to focus on the interfaces to device components that encapsulate the model implementation of each block independently.  

\subsection{Generator Models}
\begin{figure}[t]
    \centering
    \includegraphics[width=0.495\textwidth]{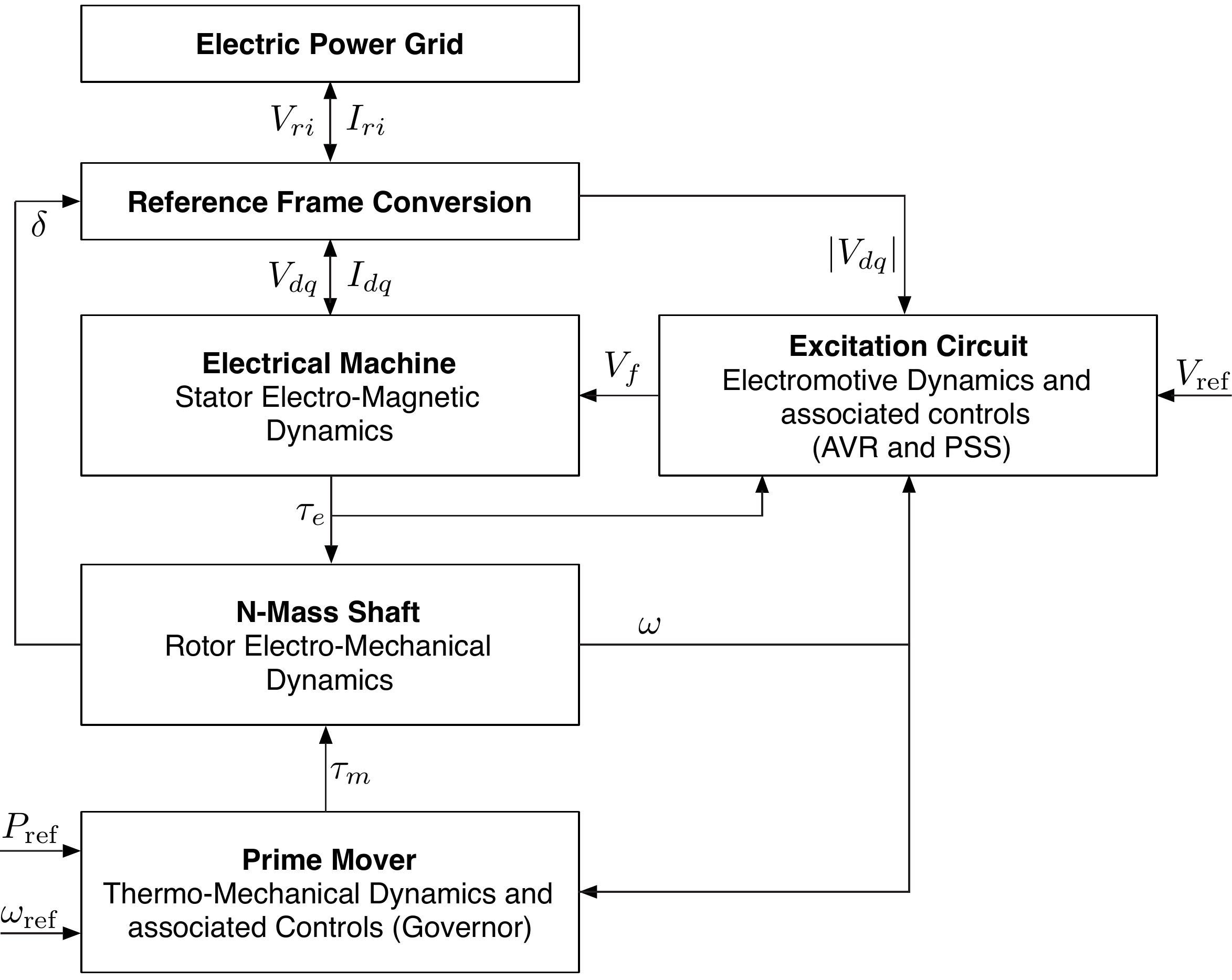}
    \caption{Generator meta-model.}
    \label{fig:gen_metamodel}
\end{figure}

Figure \ref{fig:gen_metamodel} depicts the proposed meta-model used in \texttt{LITS.jl} to describe synchronous generators. Each generator is defined by a machine model, an excitation circuit and associated controls (\ac{AVR} and \ac{PSS}), a shaft model, and a prime mover. The generator structure is designed to share network voltage and current output in the device's $dq$ \acp{SRF} $V_{dq}$, $I_{dq}$ mechanical and electrical torque $\tau_m$ $\tau_e$, frequency $\omega$, field voltage $V_f$ and angle $\delta$.

\subsection{Inverter Models}

\begin{figure}[t]
    \centering
    \includegraphics[width=0.49\textwidth]{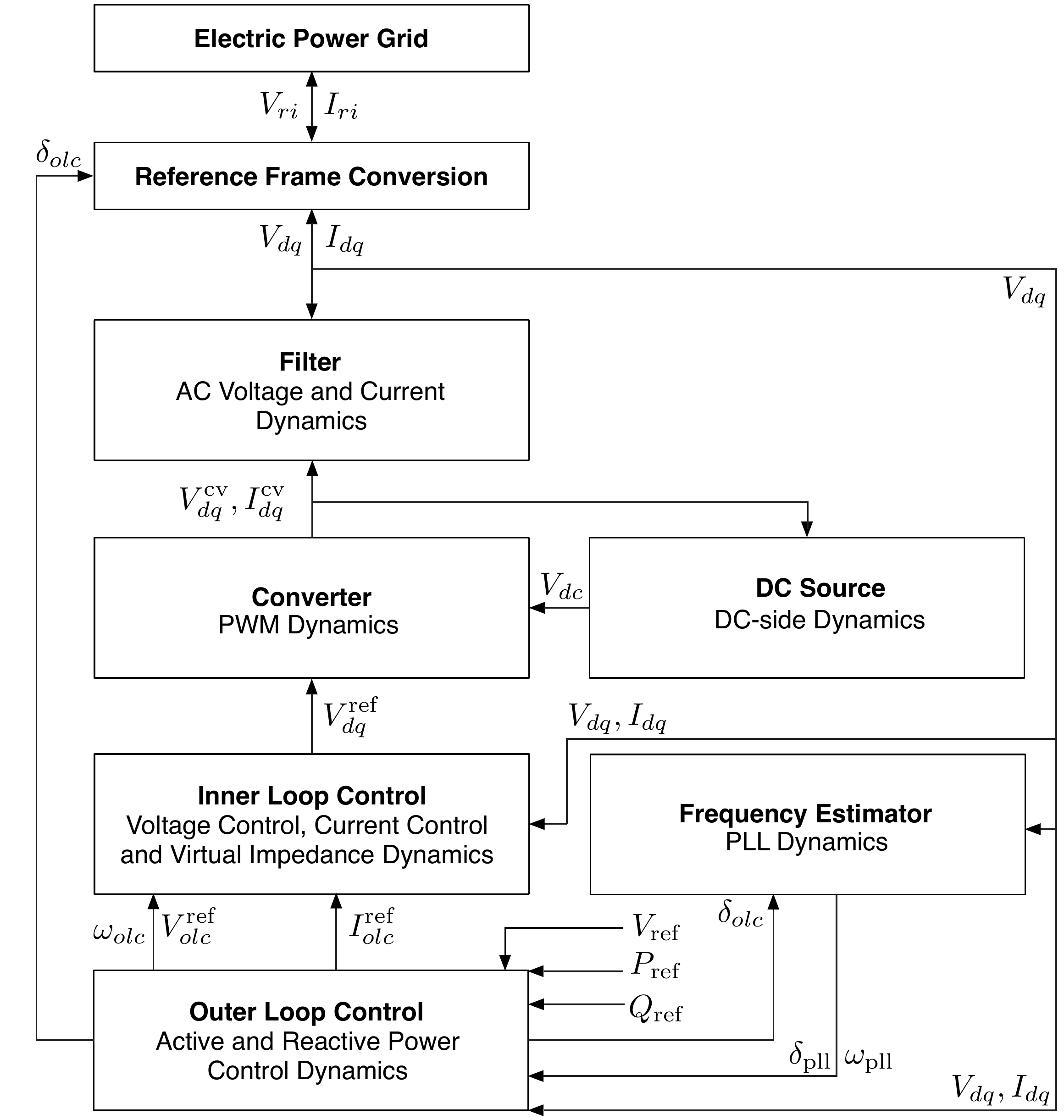}
    \caption{Inverter meta-model.}
    \label{fig:inv_metamodel}
\end{figure}

Figure \ref{fig:inv_metamodel} depicts the proposed meta-model. Each inverter is defined by a filter, converter model, inner loop control (including both voltage and current controller), outer loop control (including both active and reactive power controller), a frequency estimator (typically \ac{PLL}), and a DC source model.  

The meta-model chosen allows a high level of complexity by including the option of modeling both grid-forming and grid-feeding devices. 

\subsection{Branches}

\texttt{LITS.jl} can model AC-branch in two ways: static and dynamic. The static branch model is used to define the system's admittance matrix $\boldsymbol{Y}$ and no bus voltage differential states are included in the dynamic model. This can be used to model both lines and transformers. Dynamic branch models use the same parameters as static branch models. However, bus voltage differential equations are included in model to account for the dynamic behavior of shunt capacitance currents which are also added to the total line current balance. 

\subsection{Software Architecture}

\begin{figure}[t]
    \centering
    \includegraphics[width=0.9\columnwidth]{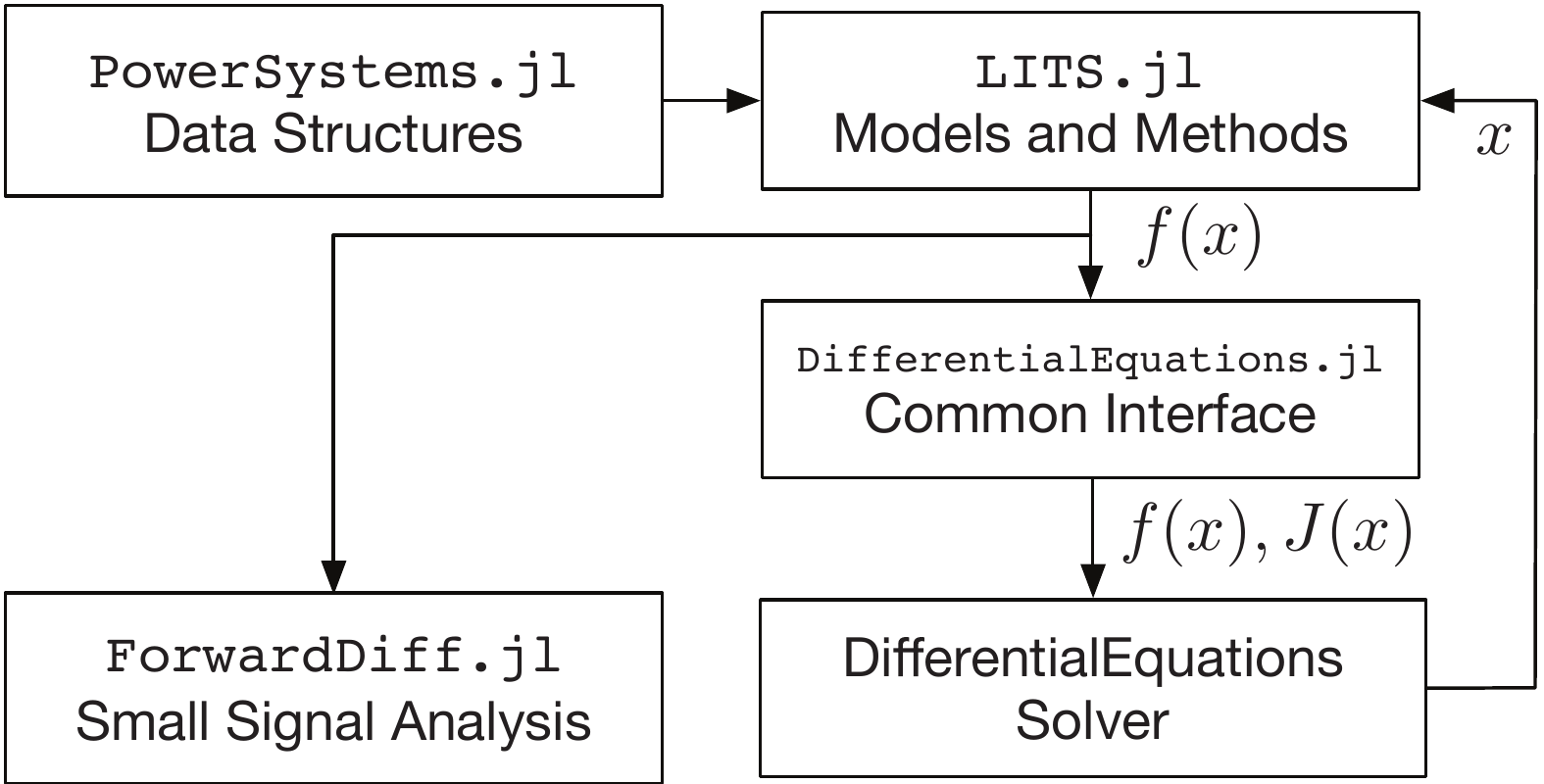}
    \caption{Information flow in \texttt{LITS.jl}.}
    \label{fig:InformationFlow}
\end{figure}

Figure \ref{fig:InformationFlow} shows the interactions between \texttt{LITS.jl}, DifferentialEquations.jl and the integrators. 
The architecture of \texttt{LITS.jl}  is such that the power system models are all self-contained and return the model function evaluations. The Jacobian is calculated through DifferentialEquations.jl's common-interface enabling the use of any solver available in Julia. Considering that the resulting models are \ac{DAE}, the implementation focuses on the use of implicit solvers, in particular SUNDIALS \cite{hindmarsh2005sundials} since it has exceptional features applicable to large models — for instance, interfacing with distributed linear-solvers and GPU arrays. 

The data model is critical to enable fast model solution times given that most of the allocating computations happen before the simulation run-time. All the model functions in \texttt{LITS.jl} are implemented as in-place updating and avoid allocations at each iteration of the integrator to reduce memory usage. In the system construction process, we generate an internal index with the location of the states relevant to each device and achieve efficient use of memory.  The information necessary for the pre-indexing exists already in the data model. The index is stored in a 2-level nested dictionary. The device name is the first level key, and the state is the second level key. As a result, whenever the function iterates over the vector containing the entire state space, only the relevant portions are accessed.  

Pre-indexing and internal variable sharing enable fast and memory-efficient evaluations of the model function and its Jacobian. In conjunction with the fact that Julia code is a Just-In-Time compiling language, \texttt{LITS.jl} achieves exceptional computational performance with reduced software development overhead when compared to low-level languages. 

At the device modeling layer, component instances share local information internal to each device through port variables. The port variables enable efficient information passing between component modeling functions. 

\subsection{Perturbations}

The DifferentialEquations.jl common interface simplifies the implementation of perturbations such as line faults or step changes. Through the use of callbacks into the solvers, it is possible to control the solution flow and intermediate initializations without requiring simulation re-starts and other heuristics typically used in power systems dynamic modeling. The built-in features and the design of perturbation objects in \texttt{LITS.jl} to internally define callbacks in the solver allows further flexibility in the type of perturbations that can be analyzed.

\subsection{Small Signal Analysis}

In order to estimate the eigenvalues and perform Small Signal Stability Analysis, the complete dynamic model containing all states from dynamic components $x$ and algebraic variables $y$ is used to define $g(y,x)$ as the vector of algebraic equations and $f(y,x,p)$ as the vector of differential equations of the entire system. With that, the non-linear differential algebraic system of equations can be written as:
\begin{align}
\left[\begin{array}{c}
 0 \\
  \dot{x}
  \end{array}\right] = \left[\begin{array}{c}
  g(y,x,p) \\
   f(y,x,p) \end{array}\right]
\end{align}
For Small Signal Stability Analysis, we are interested in the stability around an equilibrium point $y_{eq},x_{eq}$ that satisfies $\dot{x} = 0$ or equivalently $f(y_{eq},x_{eq}) = 0$, while satisfying $g(y_{eq}, x_{eq}) = 0$. To do that we use a first order approximation:
\begin{align}
\left[\begin{array}{c}
 0 \\
  \Delta\dot{x}
  \end{array}\right] = \underbrace{\left[\begin{array}{c}
  g(y_{eq},x_{eq}) \\
   f(y_{eq},x_{eq}) \end{array}\right]}_{ = 0}
 + J[y_{eq}, x_{eq}] \left[\begin{array}{c}
 \Delta y \\
  \Delta x
  \end{array}\right]
  \end{align}

For small signal analyses, we are interested in the stability of the differential states, while still considering that those need to evolve in the manifold defined by the linearized algebraic equations. Note that the Jacobian matrix can be split in four blocks depending on the specific variables we are taking the partial derivatives:
\begin{align}
J[y_{eq}, x_{eq}] =
\left[\begin{array}{cc}
 g_y & g_x \\
 f_y & f_x \\
  \end{array}\right]
\end{align}

\noindent Assuming that $g_y$ is not singular \cite{milano2010power} we can eliminate the algebraic variables to obtain the reduced Jacobian:
\begin{align}
J_{\text{red}} = f_x - f_y g_y^{-1} g_x
\end{align}
that defines our reduced system for the differential variables
\begin{align}
\Delta \dot{x} = J_{\text{red}} \Delta x
\end{align}
on which we can compute its eigenvalues to analyze local stability. To do so, \texttt{LITS.jl} computes the equilibrium point $y_{eq}, x_{eq}$ by solving the non-linear system of equations in equilibrium, and via automatic differentiation by using the package \texttt{ForwardDiff.jl}, it computes the Jacobian of the non-linear algebraic system of equations at that equilibrium point. \texttt{LITS.jl} handles the resulting Jacobian and reports the reduced Jacobian and the corresponding eigenvalues and eigenvectors.

\section{Models \label{sec:modeling}}

\texttt{LITS.jl} is based on a standard current injection model as defined in \cite{milano2010power}. The numerical advantages of current injection models outweigh the complexities of implementing constant power loads for longer-term transient stability analysis. The network is defined in a \ac{SRF}, named the RI (real-imaginary) reference frame, rotating at the constant base frequency $\Omega_b$, while each device is modeled in its own $dq$ \ac{SRF}. 

\texttt{LITS.jl} also supports several static injection devices that can inject or withdraw current from a bus instantaneously. This category includes loads and voltage and current sources. All other dynamic injections devices define local differential equations for the state of each component. 

\texttt{LITS.jl} internally tracks the current-injection balances at the nodal level from all the devices on the system. Based on the buses and branches information, the system constructor computes the admittance matrix $\boldsymbol{Y}$ assuming nominal frequency, and this is used for static branch modeling. The algebraic equations for the static portions of the network are as follows:

\subsection{Network Balance Model}

\texttt{LITS.jl} internally tracks the current-injection balances at the nodal level from all the devices on the system. Based on the buses and branches information, the system constructor computes the admittance matrix $\boldsymbol{Y}$ assuming nominal frequency and this is used for static branch modeling. The algebraic equations for the static portions of the network are as follows:
\begin{align}
    0 = \boldsymbol{i}(\boldsymbol{x},\boldsymbol{v}) - \boldsymbol{Y} \boldsymbol{v} \label{eq:current-inj}
\end{align}
where $\boldsymbol{i}$ is the vector of the sum of complex current injections from devices, $\boldsymbol{x}$ is the vector of states and $\boldsymbol{v}$ is the vector of complex bus voltages. Equations \eqref{eq:current-inj} connect all the port variables, i.e., currents, defined for each injection device. Components that contribute to \eqref{eq:current-inj} by modifying the current $\boldsymbol{i}$ or the admittance matrix $\boldsymbol{Y}$,  are (i) static injection devices, (ii) dynamic injection devices, (iii) algebraic network branches and (iv) dynamic network branches.

\subsection{Available Generator Component Models}

\begin{figure}[t]
    \centering
    \includegraphics[width=0.495\textwidth]{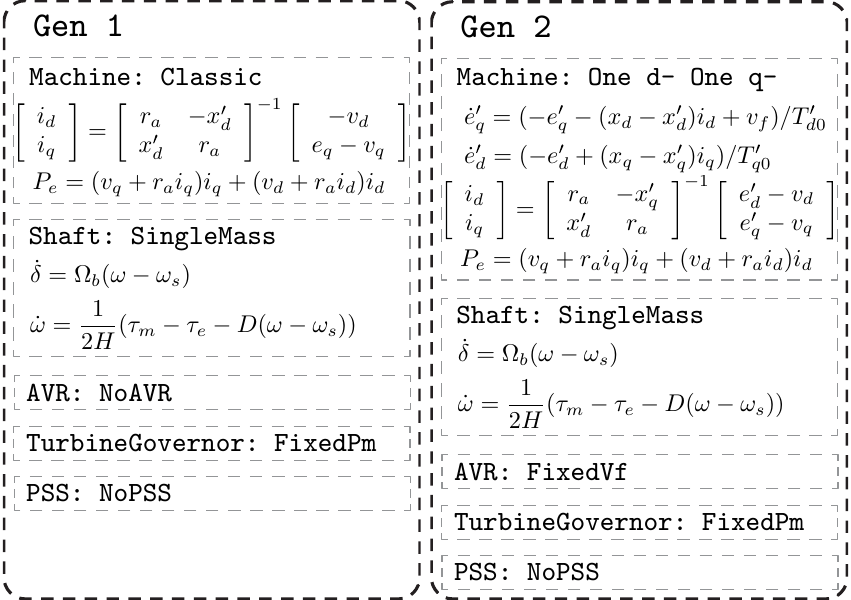}
    \caption{Mix and match different components for generators and their respective DAEs.}
    \label{fig:gen_dae_examples}
\end{figure}

Currently, for synchronous generators, the software supports:
 \begin{itemize}
\item{Electrical Machines:} The classical model (II-order model), one d- one q- axis model (IV-order model), Anderson-Fouad model (VI and VIII-order model), Marconato model (VI and VIII-order model), and stator-rotor fluxes model are already implemented \cite{milano2010power, kundur1994power}. 
\item{AVR:}
Simplified DC and AC AVR (namely Type I and Type II) are already implemented \cite{milano2010power}. In addition, fixed voltage $V_f$ and no-AVR, are also included.
\item{PSS:}
Models already implemented consider a simplified-PSS, namely a rotor velocity and electrical power droop for $V_f$, used as an extra signal for the AVR. Type I and Type II proposed in \cite{milano2010power} are also implemented.
\item{Shafts:}
A standard 1-mass shaft (rotor mass) is implemented. A 5-mass shaft (modeled as a spring-mass system) for rotor mass, high-pressure, mid-pressure, low-pressure turbines, and excitation mass, is also included \cite{milano2010power}.
\item{Prime movers:}
Type I and Type II turbine governors \cite{milano2010power} are implemented. In addition, a fixed mechanical torque $\tau_m$ is included.
\end{itemize}

Figure \ref{fig:gen_dae_examples} provides an example of the modularity capabilities when defining a generator model. As depicted, it is possible to exchange generator components to study the effects of different complexity levels. In this particular case, it is possible to exchange the generator's machine model between the classic algebraic representation in the two-state machine with the 2-state transient model of the 4-state machine. This flexibility is key to studying under-explored and emerging challenges in low-inertia systems. For example, including multi-mass shaft dynamics can be used to study torsional interaction between generators and \ac{CIG} power sources and their controllers, an issue that has been reported in systems with HVDC lines and FACTS \cite{kundur1994power}. In addition, it enables the analysis of controller architecture upgrades to enhance the resilience of the system as the adoption of \ac{CIG} sources increases. 

Details about the specific model equations and implementation can be found in the documentation\footnote{https://energy-mac.github.io/LITS.jl/}.

\subsection{Available Inverter Component Models}

The modularity of the inverter meta-model, as depicted in Fig. \ref{fig:inv_metamodel}, allows users to easily substitute any component with custom ones only requiring the implementation of the port variables -- for instance, a new \ac{PLL} for frequency estimation or outer control loops. Component model reuse permits rapid testing of different custom inverter configurations without requiring the complete re-implementation of the rest of the inverter components. For example, the outer loop control virtual speed $\omega_{olc}$ is a differential state if virtual inertia is considered, like in the \ac{VSM} model. However, in an grid-feeding inverter without virtual inertia, $\omega_{olc}$ is implemented as an algebraic state, equal to $\omega_{\text{pll}}$ that depends linearly on the \ac{PLL} voltage states \cite{lin2017stability}.  Additionally, inverters can operate in either grid-feeding or grid-supporting mode individually, which enables a heterogeneous mix of devices in the system.

The current modeling capabilities include several complex configurations, such as \ac{VSM} control using  \ac{PLL}-orientated \ac{SRF} as in \cite{DArco2015_EPSR122, markovic2018stability}.  

\begin{algorithm}[t]
\floatname{algorithm}{\footnotesize{\textbf{Code Block}}}
  \caption{{\footnotesize{Running an OMIB case in LITS v0.3.0 in Julia 1.3}}}
  \begin{algorithmic}
  {\scriptsize{ {\tt 
  \STATE \# Load LITS.jl and data
  \STATE \textbf{using} LITS
  \STATE \textbf{using} PowerSystems
  \STATE \textbf{using} Sundials
  \STATE \textbf{using} Plots
  \STATE \textbf{include}("data/network\_data.jl")
  \STATE
  \STATE \#\#\# Generator Data
  \STATE \# Machine
  \STATE mach = BaseMachine(0.0, \#R
  \STATE  \hspace{2.6cm} 0.2995, \#Xd\_p
  \STATE  \hspace{2.6cm} 0.7087, \#eq\_p
  \STATE  \hspace{2.6cm} 100.0);  \#MVABase
  \STATE \# Shaft
  \STATE shaft = SingleMass(3.148, \#H
  \STATE \hspace{2.6cm} 2.0); \#D
  \STATE \# AVR
  \STATE no\_avr = AVRFixed(0.0);
  \STATE \# Turbine Governor
  \STATE no\_tg = TGFixed(1.0); \#efficiency
  \STATE \# PSS
  \STATE no\_pss = PSSFixed(0.0);
  \STATE gen = \textbf{DynamicGenerator}(1, \#number
  \STATE \hspace{2.6cm} "OMIB\_Gen", \#GenName
  \STATE \hspace{2.6cm} buses[2], \# BusLocation
  \STATE \hspace{2.6cm} 1.0, \# $\omega$\_ref
  \STATE \hspace{2.6cm} 1.0, \# V\_ref,
  \STATE \hspace{2.6cm} 0.5, \# P\_ref,
  \STATE \hspace{2.6cm} 0.0, \# Q\_ref, not used with AVR
  \STATE \hspace{2.6cm} mach, \# machine
  \STATE \hspace{2.6cm} shaft, no\_avr, no\_tg, no\_pss) 
  \STATE
  \STATE \# Create system (MVABase = 100 MVA, nom frequency 60 Hz)
  \STATE  sys = System(100.0, frequency = 60.0)
  \STATE \# Add buses
  \STATE \textbf{for} (bus in buses) add\_components(sys, bus) \textbf{end}
  \STATE \# Add branches
  \STATE \textbf{for} (br in branches) add\_components(sys, br) \textbf{end} 
  \STATE \# Add loads
  \STATE \textbf{for} (load in loads) add\_components(sys, load) \textbf{end}
  \STATE \# Add infinite source
  \STATE add\_component!(sys, source)
  \STATE \# Add generator
  \STATE add\_component!(sys, gen)
  \STATE 
  \STATE \# Run simulation. Compute faulted Ybus.
  \STATE Ybus\_fault = Ybus(branches\_faulted, buses)
  \STATE \# Set a three phase fault (change in Ybus)
  \STATE Ybus\_change = ThreePhaseFault(1.0, \#change \STATE \hspace{4.2cm} Ybus\_fault) \#New YBus
  \STATE tspan = (0.0, 30.0) \#define time span
  \STATE \# Run a dynamic simulation.
  \STATE sim = \textbf{Simulation}(sys, \#system
  \STATE \hspace{2.4cm} tspan, \#time span
  \STATE \hspace{2.4cm} Ybus\_change) \#fault
  \STATE \#Obtain small signal results for initial conditions
  \STATE small\_sig = small\_signal\_analysis(sim)
  \STATE \# Run transient simulation 
  \STATE run\_simulation!(sim, Sundials.IDA(), dtmax = 0.02);
  \STATE \# Obtain results
  \STATE series = get\_state\_series(sim, ("OMIB\_Gen", :$\delta$)) \vspace{-0.1cm}
  \STATE plot(series)
  }}}
  \end{algorithmic}
\end{algorithm}

 \begin{itemize}
\item{Filters:}
LC and LCL filters are implemented to connect the converter with the electric grid.
\item{Converters:}
An average converter model is already implemented. Converter voltage output is given by $v_\text{cnv} = m v_\text{dc}$, where $m$ is the modulation signal -- an output of the inner loop control -- and $v_\text{dc}$ is the DC voltage output of the DC side.
\item{Inner-Loop Control:} Output current and voltage references first pass through a virtual impedance block and then through cascaded PI controllers, implemented in the synchronous reference frame of the inverter. This control implementation ultimately provides the $dq$ modulation commands to the converter model.
\item{Outer-Loop Control:}
We implement a decoupled active power controller and reactive power controller. The active power controller emulates a synchronous machine with an inertia constant, damping term, and frequency droop. The reactive power controller is implemented as a proportional voltage droop controller \cite{DArco2015_EPSR122}.
\item{Frequency Estimator:}
A \ac{PLL} is implemented to estimate the grid frequency based on the bus voltages using the model presented in \cite{DArco2015_EPSR122}.
\item{DC source:}
Currently the model supports any DC voltage source. However, the only available implementation is a fixed DC source voltage.
\end{itemize}

\begin{figure}[t]
    \centering
    \includegraphics[width=0.485\textwidth]{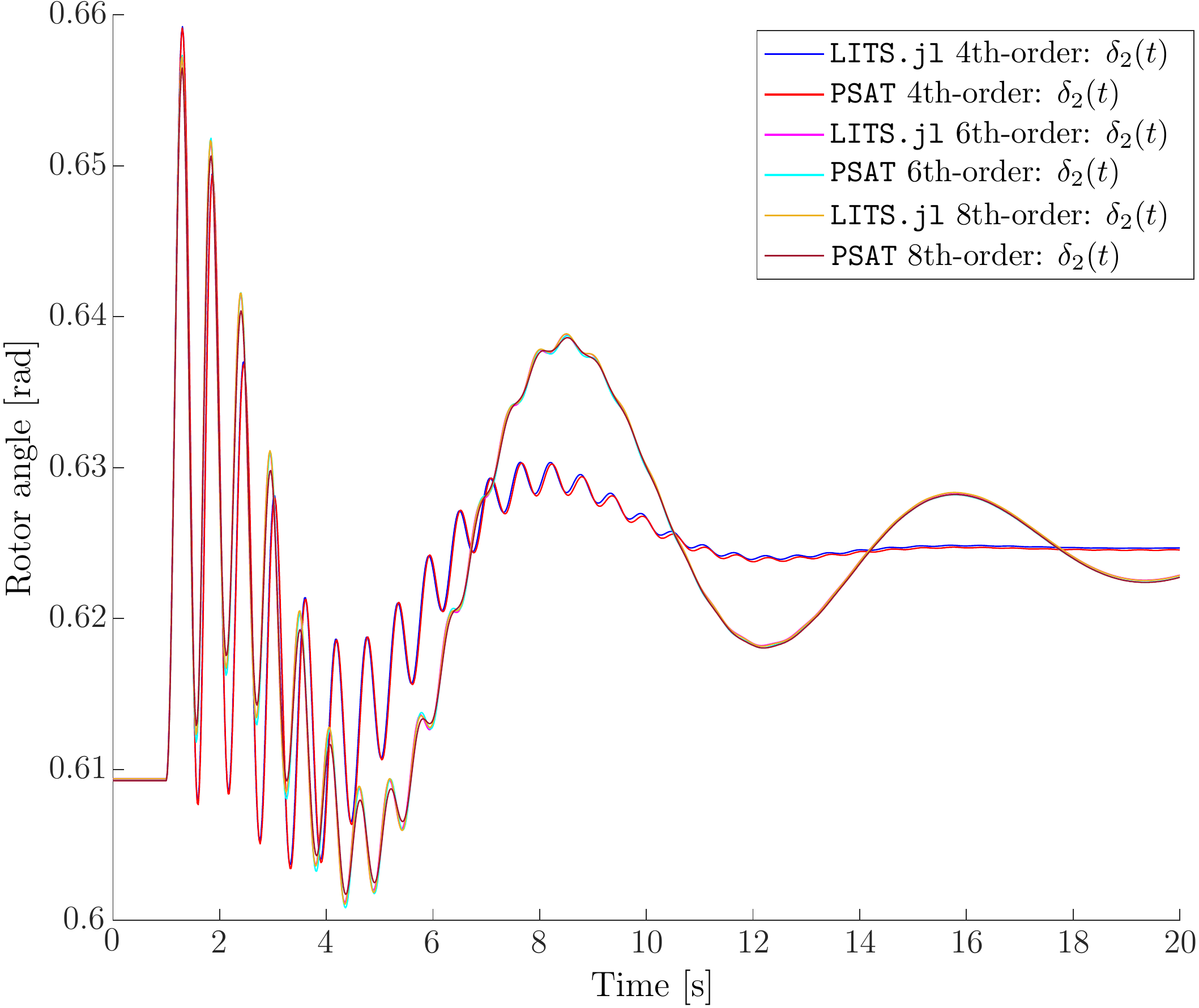}
    \caption{Rotor angle $\delta_2(t)$ evolution for different machine models in Case 1.}
    \label{fig:machine_benchmark}
\end{figure}

\subsection{Branch models}

Each dynamic branch will add three differential equations as follows:
\begin{align}
    \frac{L}{\Omega_b} \frac{di_\ell}{dt} &= (v_\text{from} - v_\text{to}) - (R+jL) i_\ell \\
     \frac{C_\text{from}}{\Omega_b} \frac{dv_\text{from}}{dt} &=  i_c^{\text{from}} - jC_\text{from}v_\text{from}   \\
      \frac{C_\text{to}}{\Omega_b} \frac{dv_\text{to}}{dt} &= i_c^{\text{to}} - jC_\text{to}v_\text{to}
\end{align}
where $i_\ell$ is the complex current going through the RL circuit and $i_c$ is the current injected due to the line shunt susceptance included in the $\Pi$-line model.

\section{Simulations \label{sec:simulations}}

Case studies and benchmarks are presented to showcase the capabilities of \texttt{LITS.jl} and validation of the implemented models. The first case study focus on synchronous generator modeling and compare the simulations to PSAT \cite{milano2005psat}.  The second case compares the inverter model with published results in \cite{DArco2015_EPSR122}. The third case showcase the effects of considering multi-mass shaft models. Finally, the fourth case analyzes the interaction between a synchronous machine and an inverter in a 3-bus system. The last case also highlight the capabilities of the proposed software to model dynamic branches selectively. 

The simulations cases use the implicit solver IDA from the suite SUNDIALS \cite{hindmarsh2005sundials}, which is available in the Julia environment. 

\subsection{Running a case}

\texttt{LITS.jl} is a registered Julia package. To install, using the Julia REPL: \texttt{] add LITS}. The source code can be found in the repository\footnote{\url{https://github.com/Energy-MAC/LITS.jl}} where the master branch contains the latest development code. Code Block 1 describes a basic simulation run using \texttt{LITS.jl} of a 2-state machine against infinite bus. This case is available in the Github repository of \texttt{LITS.jl} examples.

\subsection{Case 1: 3 buses -- Two generators against infinite bus}

The first case study analyzes a triangular 3-bus system. At bus 1, a voltage source is connected as an infinite bus. Single shaft machines are connected in buses 2 and 3. Each machine has an AVR Type I (simplified IEEE DC AVR model). The bus loads are 1.5 in buses 1 and 2 and 0.5 MW in bus 3. The simulation is developed to analyze the behavior of several machine electromagnetic models. In this particular case, we analyze the following 
\begin{enumerate}
    \item One d- One q- (4th order machine).
    \item Simplified Marconato model (6th order machine).
    \item Full Marconato model (8th order machine) \cite{marconato2002electric}.
\end{enumerate}

The results show a line trip perturbation triggered at $t=1$s. Figure \ref{fig:machine_benchmark} shows the evolution of rotor angle of the synchronous generator connected at bus 2. Results show that the responses between PSAT and \texttt{LITS.jl} are equivalent with acceptable differences in the range of numeric tolerances.

\begin{figure}[t]
    \centering
    \includegraphics[width=0.44\textwidth]{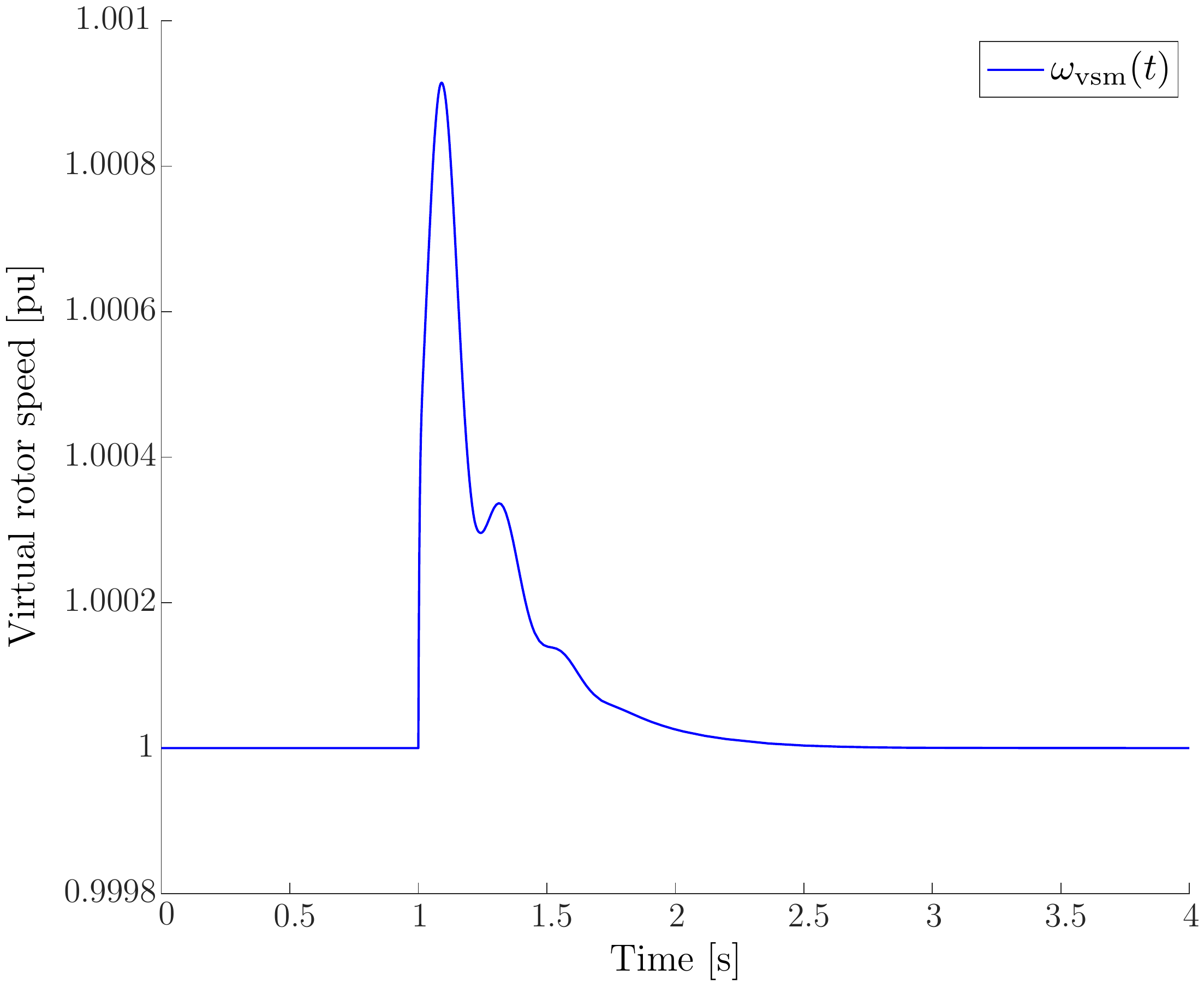}
    \caption{Virtual rotor speed $\omega_\text{vsm}(t)$ evolution for VSM in Case 2.}
    \label{fig:darco_benchmark}
\end{figure}

\subsection{Case 2: VSM against infinite bus}

The second case study is used to validate the transient behavior of a 19-states \ac{VSM} inverter connected against an infinite bus with the same setting as in \cite{DArco2015_EPSR122}. A step change in the power from 0.5 pu to 0.7 pu is introduced at $t=1$. Figure \ref{fig:darco_benchmark} depicts the effect on the \ac{VSM} angular speed due to the step change. Results show a consistent transient behavior that is showcased in Figure 9 in \cite{DArco2015_EPSR122}.

\subsection{Case 3: Multi-mass shaft modeling}

\begin{figure}[t]
\vspace{0.2cm}
    \centering
    \includegraphics[width=0.485\textwidth]{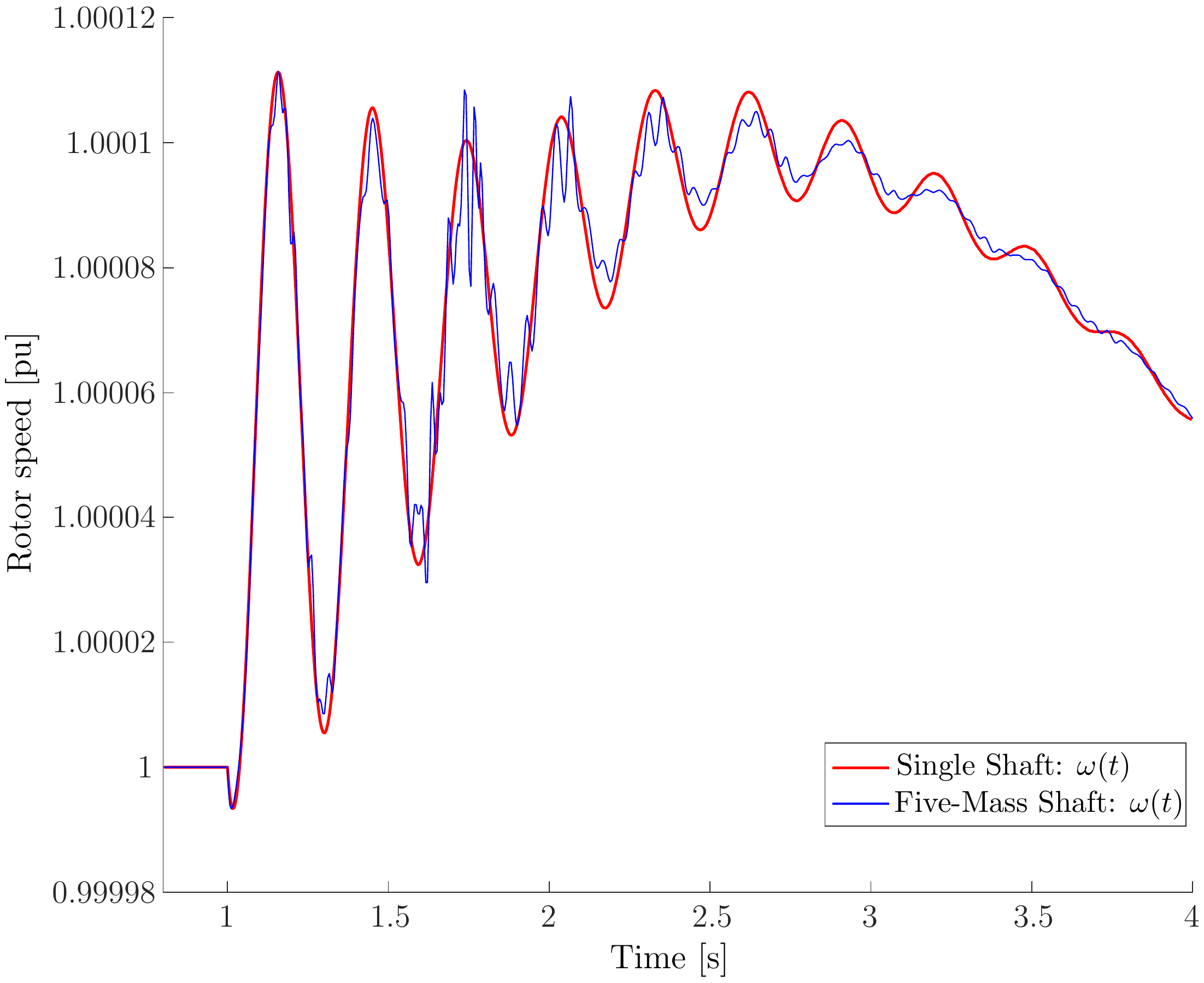}
    \caption{Rotor speed $\omega(t)$ evolution for different shaft models in Case 3.}
    \label{fig:shaft_comparison}
\end{figure}

Our third case uses the same system as in Case 1 with a One d- one q- (4th order) machine in Bus 2 and replacing the single-mass shaft model for a five-mass spring dynamics model. This is directly implemented in \texttt{LITS.jl} by modifying the generator shaft of the generator structure. Similar to Case 1, at $t=1$s, the line that connects buses 1 and 3 trips. Figure \ref{fig:shaft_comparison} depicts the effect on rotor speed $\omega(t)$ of including spring dynamics at the shaft between mass elements. Due to the presence of the infinite bus and power flow reconfiguration, rotor speed remains close to the nominal frequency, but faster oscillations can be observed in the five-mass shaft due to the additional second order dynamics.

\subsection{Case 4: \ac{VSM} against synchronous machine considering dynamic lines}

The fourth case is also based on the same system as Case with One d- one q- (4th order) machine in Bus 2 and a \ac{VSM} inverter connected in bus 3. The model also considers different line models connecting bus 2 and 3, first an algebraic one, while on a second scenario we use a dynamic line model. All remaining lines are modeled as algebraic ones in both scenarios. This is done in \texttt{LITS.jl} by simply selecting which data types of lines are static or dynamic.

\begin{figure}[t]
    \centering
    \includegraphics[width=0.485\textwidth]{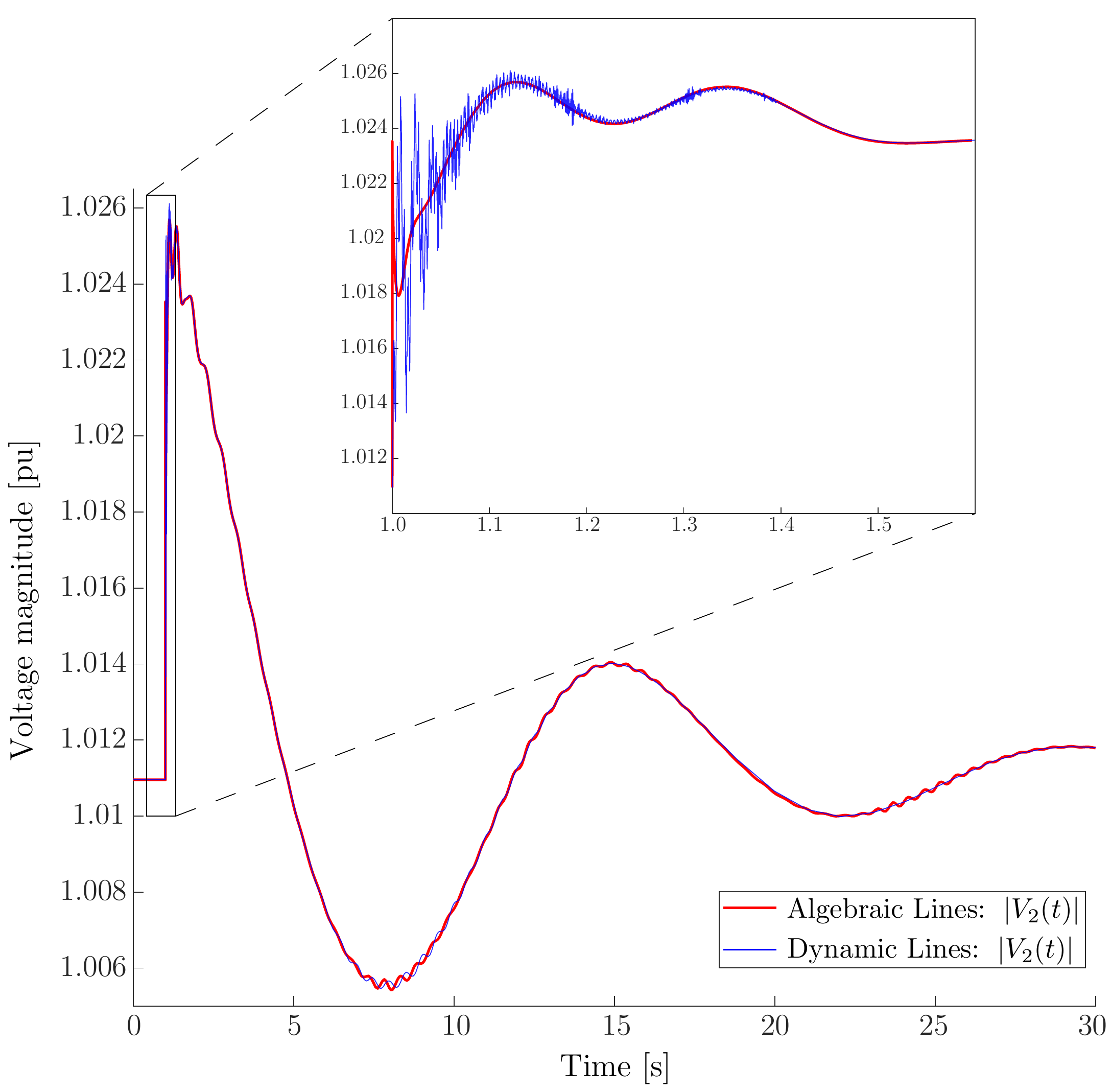}
    \caption{Voltage magnitude $|V_2(t)|$ at bus 2, for Case 4.}
    \label{fig:dynlines_voltages}
\end{figure}

At $t=1$, two of the three circuits of the line that connects buses 1 and 3 trip. Figure \ref{fig:dynlines_voltages} illustrates the evolution of the voltage magnitude at bus 2. As depicted, line dynamics allow us to model a fast behavior that is not reflected in the static lines simulation case that have effects on the transient response of different devices. As mentioned before, a key motivation for building \texttt{LITS.jl} is to enable this trade-off between model complexity and computational requirements. More complex models, such as ones including line dynamics, showcase faster behavior, but lead to increasingly stiff systems, longer computational times, and susceptibility to numerical instabilities under default solver parameters.

\section{Conclusions \label{sec:conclusion}}

This paper introduced the open-source simulation toolbox \texttt{LITS.jl}, which focuses on transient simulation of low-inertia power systems. \texttt{LITS.jl} is built on Julia scripting language and incorporates a myriad of power system devices and component models.  \texttt{LITS.jl} enables the analysis of transient responses of dynamical systems using different model complexity. The development of device meta-models for generators and inverters standardize ports and states interaction among devices and their internal components. The proposed software design enables researchers to define a new component's model within the proposed meta-model and quickly explore novel architectures and controls.

We presented several numerical experiments and benchmarks to highlight the capabilities of \texttt{LITS.jl} and the validity of the models included in the library. Case studies show that \texttt{LITS.jl} enables easier assessment of the trade-off between model complexity and computational requirements by exchanging device models and comparing simulation results under different assumptions. 

The ongoing development of  \texttt{LITS.jl} focuses on extending the available models and analytical capabilities. This includes reduced-order inverter models, FACTS device models, and HVDC and DC-side dynamics of inverters.  Finally, incorporating numerical Jacobian calculations capabilities for small-signal analysis is a future task in this project.



\bibliographystyle{IEEEtran}
\bibliography{references}
%

\end{document}